# Comparison of the Lindhard−Sørensen and Mott−Bloch corrections to the Bethe stopping formula at moderately relativistic energies[1]

P.B. Kats*[2], K.V. Halenka*, O.O. Voskresenskaya**[3]

*Brest State A.S. Pushkin University, Brest, 224016, Belarus
**Joint Institute for Nuclear Research, Dubna, Moscow Region, 141980 Russia

*The results of numerical calculating the total Mott−Bloch correction to the Bethe stopping formula and the Lindhard−Sørensen correction in the point nucleus approximation, as well as the Mott correction and the difference between the Lindhard−Sørensen and Bloch corrections, which were obtained by some rigorous and approximate methods, are compared for the ranges of a gamma factor 1 $\lesssim \gamma \lesssim$ 10 and the ion nuclear charge number 6 $\le Z \le$ 114. It is shown that the accurate calculation of the Mott−Bloch correction based on the Mott exact cross section using a method previously proposed by one of the authors gives excellent agreement between its values and the values of the Lindhard−Sørensen correction in the $\gamma$ and Z ranges under consideration. In addition, it is demonstrated that the results of stopping power calculations obtained by the two above-mentioned rigorous methods coincide with each other up to the seventh significant digit and provide the best agreement with experimental data in contrast with the results of some approximate methods, such as the methods of Ahlen, Jackson−McCarthy, etc.*

## 1 Introduction

Research on the penetration of heavy ions in a material and the material stopping power is of great applied interest in the field of materials and surface science, radiation medicine and biology, as well as for medical, nuclear and aerospace engineering (in particular, in ion-beam therapy, ion implantation, ion beam-analysis, and ion-beam modification of materials) [1, 2].

Electronic stopping of a point relativistic heavy ion in solids is described by the relativistic version of the Bethe formula [3] that is obtained the first-order Born approximation. This formula, taking into account the density effect, reads

$$-\frac{d\bar{E}}{dx} = \zeta L,\ \ L = L_0 = \ln\left(\frac{E_m}{I}\right) - \beta^2 - \frac{\delta}{2},$$

$$\zeta = 4\pi r^2 mc^2 \cdot N_e \cdot \left(\frac{Z}{\beta}\right)^2 = 4\pi r^2 mc^2 \cdot N_A \rho \frac{Z'}{A} \cdot \left(\frac{Z}{\beta}\right)^2,\ \ E_m = \frac{2mc^2\beta^2}{1-\beta^2},$$

or, in units MeV $g^{-1}cm^2$, it can be rewritten as follows:

$$-\frac{d\bar{E}}{\rho dx} = \tilde{\zeta} L,\ \tilde{\zeta} = 4\pi r^2 mc^2 \cdot \tilde{N}_e \cdot \left(\frac{Z}{\beta}\right)^2 = 0.307075 \frac{Z'}{A}\left(\frac{Z}{\beta}\right)^2. \quad (1)$$

---

[1] arXiv:2008.04819v2
[2] Email: aid@tut.by
[3] Email: voskr@jinr.ru

In these equations, $x$ denotes the distance traveled by a particle; $L$ is the so-called 'stopping number' and $L_0$ denotes the 'Bethe logarithm'; $E_m$ is the maximum transferrable energy to an electron of mass $m$ and classical radius $r = e^2/(mc^2)$ in a collision with the particle of velocity $\beta c$; $I$ is the effective ionization potential of the absorber atoms; $Z$ is the charge number of incident nucleus; $\delta/2$ describes the density effect correction of Fermi; and $N_e$ is the electron density of a material that is either measured in electrons/g ($\tilde{N}_e = N_A Z'/A$) or in electrons/cm$^3$ ($N_e = N_A \rho Z'/A$), where $\rho$ is density of a material in g cm$^{-3}$, $N_A$ denotes the Avogadro number, $Z'$ and $A$ refer to the atomic number and weight of the absorber [4, 5].

The above expressions are applicable if $Z\alpha / \beta << 1$, where $\alpha$ is the fine-structure constant. If this condition is not satisfied, the Bloch correction $\Delta L_B$ [6] to the $L_0$ and the Mott correction $\Delta L_M$ based on the Mott exact cross section [7] are also introduced:

$$\Delta L_B = \psi(1) - \mathrm{Re}\,\psi(1 + iZ\alpha / \beta)$$

with the digamma function $\psi$ and

$$\Delta L_M = \frac{\tilde{N}_e}{\tilde{\zeta}} \int_{\varepsilon}^{E_m} E\left[\left(\frac{d\sigma}{dE}\right)_M - \left(\frac{d\sigma}{dE}\right)_{FB}\right] dE. \qquad (2)$$

Here, $\varepsilon$ is some energy above which the atomic electron binding energy may be neglected, and $(d\sigma/dE)_{M(FB)}$ are, respectively, the Mott and Born expressions for the scattering cross section of electrons on nuclei. Switching in the expression (2) from integration over the energy $E$ transferred to an electron to integration over the center-of-mass scattering angle $\theta$, we can rewrite (2) in the form

$$\Delta L_M = 2\pi \frac{\tilde{N}_e E_m}{\tilde{\zeta}} \int_{\theta_0}^{\pi} \left[\left(\frac{d\sigma(\theta)}{d\Omega}\right)_M - \left(\frac{d\sigma(\theta)}{d\Omega}\right)_{FB}\right] \sin^2\left(\frac{\theta}{2}\right) \sin\theta d\theta, \qquad (3)$$

where $\theta_0$ denotes the scattering angle corresponding to $\varepsilon$ and $\Omega$ is the scattering cross section solid angle.

In the range $\gamma \lesssim 10$, the stopping power is well-described by (1) including the stopping number $L = L_0 + \Delta L$ with the sum of the above corrections,

$$\Delta L = \Delta L_{MB} = \Delta L_M + \Delta L_B. \qquad (4)$$

The importance of this total 'Mott−Bloch correction' was shown, in particular, for the interpretation of the experiment at the GSI/SIS accelerator ($\gamma \sim 2$) [8] and other experiments (e.q. [9-11]).

The Mott correction was first observed experimentally by Tarle and Solarz [12] and later measured with greater precision by Salamon et al. [13]. It was first calculated by Eby and Morgan [14, 15] by numerical integration of (2) for several values of $Z$ and $\beta$. These calculations demonstrated the significance of Mott's correction to the Bethe−Bloch formula for incident nuclei with $Z \geq 20$.

Since the expressions (2) and (3) for $\Delta L_M$ are extremely inconvenient for practical application, the analytical expressions for $\Delta L_M$ in the second[4] and third order Born approximations were also proposed [15] based on the relevant approximate McKinley−Feshbach and Johnson−Weber−Mullin results for the Mott exact cross section [17, 18]. A closed third order Born expression for $\Delta L_M$ was also obtained by Ahlen [19], and several other approximate expressions were proposed for this correction (see e.g. [20]). The drawback of these approximate expressions is their restricted range of application, roughly estimated by the relation $Z/\beta \leq 100$, and the essentially uncertain accuracy. Moreover, the incorrect threshold (in the limit $\beta \to 0$) behavior of these expressions precludes their use for calculating the total ranges of relativistic heavy ions in matter. Therefore, obtaining convenient rigorous expressions for corrections to the Bethe logarithm is very important.

In 1996 it was shown that computing the integrals (2), (3) can be reduced to a summing the fast converging infinite series whose terms are bilinear in the Mott partial amplitudes and a question was raised regarding the choice of an efficient method for numerical summation of these series [21]. In the same year, Lindhard and Sørensen proposed a correction to the Bethe equation, taking into account a finite size of the projectile nucleus at ultrarelativistic energies ($\gamma \gtrsim 10$) [22] and their prediction of the finite nuclear size effect was confirmed at the CERN/SPS accelerator with the 160 GeV/u Pb beam ($\gamma = 168$) [23][5]. As in the previous period approximate methods for calculating the Mott correction became widespread (the Jackson−McCurthy, Allen methods and others), the Mott correction began to be identified with its approximations and an opinion began to form about the 'approximate nature' of this correction, as well as about replacing the total Mott−Bloch correction with 'more precise' correction of Lindhard and Sørensen [22, 25, 26].

In this work, we will carry out a numerical investigation which shows that at moderately relativistic energies ($\gamma \lesssim 10$), when a projectile can be considered as a point-like particle, the method based on calculating the Mott-exact cross section and the Lindhard−Sørensen method give completely coinciding results, while the results of approximate methods for calculating the Mott corrections and stopping power differ significantly from these results. The outline of this paper is as follows. We first consider the formulas that used later in the calculation of the corrections to the Bethe sopping power. Then we present numerical results for these corrections and the stopping power based on them. Finally, we short summarize our findings.

This paper is devoted to the memory of Alexander Tarasov, a remarkable scientist and person [27][6] who owns a decisive contribution to the work [21].

---

[4] This result has been previously obtained by Jackson and McCurthy [16].
[5] Ref. [24] presents Jens Lindhardt's contribution to the stopping power theory.
[6] A collection of selected works by Alexander Tarasov, which also contains essays on his life and scientific activity.

## 2 Basic formulae

The Mott corrections were calculated by us with the aid of the method [21] and using the approximations of Jackson and McCurthy (second Born approximation) [16], Morgan and Eby (third-order Born approximation) [15], Ahlen [19], and Matveev [20].

The second-order Born approximation to the Mott correction obtained by Jackson and McCurthy [16] and independently by Morgan and Eby [15], based on the approximate McKinley–Feshbach results for the Mott-exact cross section $(d\sigma/dE)_{MMF}$ [17], reads

$$\Delta L_{MJM} = \frac{\tilde{N}_e}{\tilde{\zeta}} \int_0^{E_m} E\left[\left(\frac{d\sigma}{dE}\right)_{MMF} - \left(\frac{d\sigma}{dE}\right)_{FB}\right] dE = \frac{1}{2}\pi\alpha\beta Z. \quad (5)$$

From the approximate Johnson–Weber–Mullin results for the Mott-exact cross section $(d\sigma/dE)_{MJWM}$ [18], Morgan and Eby [15] obtained the following closed form for the third-order Born approximation to the $\Delta L_M$:

$$\Delta L_{MME} = \frac{\tilde{N}_e}{\tilde{\zeta}} \int_0^{E_m} E\left[\left(\frac{d\sigma}{dE}\right)_{MJWM} - \left(\frac{d\sigma}{dE}\right)_{FB}\right] dE = $$
$$= \frac{1}{2}\left\{\pi\alpha\beta Z + (\alpha Z)^2\left[\pi^2/3 + 1 + \beta^2(3/4 - \ln 2) + 0.5(\zeta(3) - 3)\right]\right\}. \quad (6)$$

Here $\zeta(3)$ is the Riemann Zeta function.

Ahlen [19] has taken advantage of the $Z^7$ expansion derived by Curr [28] for the Mott cross section to obtain an analytical expression for the Mott correction. The form recommended by Ahlen for $\Delta L_M$ is as follows:

$$\Delta L_{MA} = \frac{1}{2}\eta\beta^2\{[1.725 + 0.52\pi\cos\chi] + \eta[3.246 - 0.451\beta^2] +$$
$$+\eta^2[0.987 + 1.552\beta^2] + \eta^3[-2.696 + \beta^2(4.569 - 0.494\beta^2)] +$$
$$+\eta^4[-1.170 + \beta^2(0.222 + 1.254\beta^2)]\},$$
$$\cos\chi = \mathrm{Re}\frac{\Gamma(1/2 - i\eta)\Gamma(1 + i\eta)}{\Gamma(1/2 + i\eta)\Gamma(1 - i\eta)}, \; \eta = Z\alpha/\beta. \quad (7)$$

The function $\cos\chi$ is defined by Doggett and Spencer [29] and is tabulated in [30] for various values of $\eta$.

An another convenient approximation for $\Delta L_M$ (with restriction $Z \le 92$ and $\gamma \le 10$) is proposed by Matveev [20]:

$$\Delta L_{MMT} = \ln\left[f(Z,\beta)\right],$$
$$f(Z,\beta) = 1 + \left\{0.222592\beta - 0.042948\beta^2 + \left(0.6016 + 5.15289\beta - 3.73293\beta^2\right)Z\alpha\right.$$
$$\left. -\left(0.52308 + 5.71287\beta - 8.11358\beta^2\right)(Z\alpha)^2\right\}^2. \quad (8)$$

The problem of calculating the Mott correction to all orders in $Z\alpha$ was solved by authors of [21] for the limit $\varepsilon \to 0$,

$$\Delta L_{MVSTT} = \frac{mc^2\beta^2}{4\pi(Ze^2)^2} \lim_{\varepsilon\to 0} \int_{\varepsilon}^{E_m} E\left[\left(\frac{d\sigma}{dE}\right)_M - \left(\frac{d\sigma}{dE}\right)_{FB}\right] dE,$$

where $\Delta L_M$ was expressed in terms of the rapidly converging series,

$$\Delta L_{MVSTT} = \frac{2}{\eta^2} \sum_{k=0}^{\infty} \frac{k(k+1)+\xi^2}{2k+1} \left(\left|F_M^{(k)}\right|^2 - \left|F_Z^{(k)}\right|^2\right),$$

$$F_M^{(k)} = \frac{i}{2}(-1)^k \left[kC_M^{(k)} + (k+1)C_M^{(k+1)}\right],\ F_Z^{(k)} = \frac{i}{2}(-1)^k \left[kC_Z^{(k)} + (k+1)C_Z^{(k+1)}\right],$$

$$C_Z^{(k)} = e^{-i\pi k} \frac{\Gamma(k-i\eta)}{\Gamma(k+1+i\eta)},\ C_M^{(k)} = e^{-i\pi\rho_k} \frac{\Gamma(\rho_k - i\eta)}{\Gamma(\rho_k+1+i\eta)},\ \eta = \frac{Z\alpha}{\beta},\ \xi = \frac{\eta}{\gamma},\ \rho_k = \sqrt{k^2 - Z^2\alpha^2}. \tag{9}$$

The Lindhard−Sørensen correction was derived by authors [22] using the exact solution to the Dirac equation with spherically symmetric potential. For pointlike nuclei, it can be represented as [24]

$$\Delta L_{LS} = \frac{\beta^2}{2} + \frac{1}{\eta^2} \sum_{k=1}^{\infty} k\left[\frac{k-1}{2k-1}\sin^2(\delta_k - \delta_{k-1}) + \frac{k+1}{2k+1}\sin^2(\delta_{-k} - \delta_{-k-1}) + \frac{\eta^2}{(4k^2-1)(\gamma^2k^2+\eta^2)} - \frac{\eta^2}{k^2}\right],$$

$$\delta_k = \varphi_k - \arg\Gamma(\rho_k + 1 + i\eta) + \frac{\pi}{2}(l - \rho_k),\ e^{2i\varphi_k} = \frac{k - i\xi}{\rho_k - i\eta},\ l = \begin{cases} k, & k > 0, \\ -k-1, & k < 0. \end{cases} \tag{10}$$

Here, $\delta_k$ is the Coulomb phase shifts and $\gamma$ is identical to the usual Lorentz factor $\gamma = (1-\beta^2)^{-1/2}$ . The effect of finite nuclear size appears as a modification to the Coulomb phase shifts $\delta_k$ in (11). If we represent the Bloch correction as a series [22],

$$\Delta L_B = \sum_{k=1}^{\infty} \left(\frac{k}{k^2+\eta^2} - \frac{1}{k}\right), \tag{11}$$

we can write the difference between the Lindhard−Sørensen and Bloch corrections as follows:

$$\Delta L_{LS-B} = \frac{\beta^2}{2} + \frac{1}{\eta^2} \sum_{k=1}^{\infty} k\left[\frac{k-1}{2k-1}\sin^2(\delta_k - \delta_{k-1}) + \frac{k+1}{2k+1}\sin^2(\delta_{-k} - \delta_{-k-1}) + \frac{\eta^2}{(4k^2-1)(\gamma^2k^2+\eta^2)} - \frac{\eta^2}{k^2+\eta^2}\right]. \tag{12}$$

## 3 Numerical results for the Lindhard−Sørensen and Mott−Bloch corrections

The numerical values of the $\Delta_{LS}$ and $\Delta L_{MB}$ corrections were found by us by the methods [22] and [21] over the $Z$ and $\beta$ ranges $6 \le Z \le 114$ and $0.150 \le \beta \le 0.995$ using the Wolfram Mathematica computing system. These results were also compared with the total Mott−Bloch correction computed in [15] by numerically integrating the Mott cross section (see Table 1 and Figure 1).

As can be seen, there is a remarkable agreement between the Lindhard−Sørensen correction and the total Mott−Bloch correction obtained by the method [21]. In both cases, the summation was carried out up to $k$ = 5000. Since both results are based on the solution of the Dirac equation in the Coulomb field, this agreement is explainable.[7]

Table 1: Lindhard−Sørensen ($\Delta L_{LS}$) correction in the point nucleus approximation and the Mott−Bloch ($\Delta L_{MB}$) correction obtained by the VSTT, MT, and ME methods over the $Z$ and $\beta$ ranges $6 \le Z \le 114$ and $0.85 \le \beta \le 0.99$.

| β/Z | 6 | 12 | 26 | 36 | 52 |
|---|---|---|---|---|---|
| 0.85 | $\Delta L_{LS}$ =0.059<br>$\Delta L_{MBVSTT}$ =0.059<br>$\Delta L_{MBMT}$ =0.061<br>$\Delta L_{MBME}$ =0.065 | $\Delta L_{LS}$ =0.120<br>$\Delta L_{MBVSTT}$ =0.120<br>$\Delta L_{MBMT}$ =0.110<br>$\Delta L_{MBME}$ =0.125 | $\Delta L_{LS}$ =0.267<br>$\Delta L_{MBVSTT}$ =0.267<br>$\Delta L_{MBMT}$ =0.258<br>$\Delta L_{MBME}$ =0.269 | $\Delta L_{LS}$ =0.377<br>$\Delta L_{MBVSTT}$ =0.377<br>$\Delta L_{MBMT}$ =0.380<br>$\Delta L_{MBME}$ =0.379 | $\Delta L_{LS}$ =0.562<br>$\Delta L_{MBVSTT}$ =0.562<br>$\Delta L_{MBMT}$ =0.583<br>$\Delta L_{MBME}$ =0.564 |
| 0.90 | $\Delta L_{LS}$ =0.063<br>$\Delta L_{MBVSTT}$ =0.063<br>$\Delta L_{MBMT}$ =0.065<br>$\Delta L_{MBME}$ =0.069 | $\Delta L_{LS}$ =0.128<br>$\Delta L_{MBVSTT}$ =0.128<br>$\Delta L_{MBMT}$ =0.111<br>$\Delta L_{MBME}$ =0.125 | $\Delta L_{LS}$ =0.288<br>$\Delta L_{MBVSTT}$ =0.288<br>$\Delta L_{MBMT}$ =0.273<br>$\Delta L_{MBME}$ =0.293 | $\Delta L_{LS}$ =0.411<br>$\Delta L_{MBVSTT}$ =0.411<br>$\Delta L_{MBMT}$ =0.409<br>$\Delta L_{MBME}$ =0.413 | $\Delta L_{LS}$ =0.621<br>$\Delta L_{MBVSTT}$ =0.621<br>$\Delta L_{MBMT}$ =0.644<br>$\Delta L_{MBME}$ =0.622 |
| 0.95 | $\Delta L_{LS}$ =0.067<br>$\Delta L_{MBVSTT}$ =0.067<br>$\Delta L_{MBMT}$ =0.067<br>$\Delta L_{MBME}$ =0.073 | $\Delta L_{LS}$ =0.136<br>$\Delta L_{MBVSTT}$ =0.136<br>$\Delta L_{MBMT}$ =0.118<br>$\Delta L_{MBME}$ =0.143 | $\Delta L_{LS}$ =0.309<br>$\Delta L_{MBVSTT}$ =0.309<br>$\Delta L_{MBMT}$ =0.284<br>$\Delta L_{MBME}$ =0.313 | $\Delta L_{LS}$ =0.443<br>$\Delta L_{MBVSTT}$ =0.443<br>$\Delta L_{MBMT}$ =0.434<br>$\Delta L_{MBME}$ =0.443 | $\Delta L_{LS}$ =0.676<br>$\Delta L_{MBVSTT}$ =0.676<br>$\Delta L_{MBMT}$ =0.701<br>$\Delta L_{MBME}$ =0.675 |
| 0.97 | $\Delta L_{LS}$ =0.068<br>$\Delta L_{MBVSTT}$ =0.068<br>$\Delta L_{MBMT}$ =0.068<br>$\Delta L_{MBME}$ =0.076 | $\Delta L_{LS}$ =0.139<br>$\Delta L_{MBVSTT}$ =0.139<br>$\Delta L_{MBMT}$ =0.119<br>$\Delta L_{MBME}$ =0.146 | $\Delta L_{LS}$ =0.317<br>$\Delta L_{MBVSTT}$ =0.317<br>$\Delta L_{MBMT}$ =0.288<br>$\Delta L_{MBME}$ =0.321 | $\Delta L_{LS}$ =0.455<br>$\Delta L_{MBVSTT}$ =0.455<br>$\Delta L_{MBMT}$ =0.443<br>$\Delta L_{MBME}$ =0.457 | $\Delta L_{LS}$ =0.698<br>$\Delta L_{MBVSTT}$ =0.698<br>$\Delta L_{MBMT}$ =0.723<br>$\Delta L_{MBME}$ =0.705 |
| 0.99 | $\Delta L_{LS}$ =0.070<br>$\Delta L_{MBVSTT}$ =0.070<br>$\Delta L_{MBMT}$ =0.069<br>$\Delta L_{MBME}$ =0.112 | $\Delta L_{LS}$ =0.142<br>$\Delta L_{MBVSTT}$ =0.142<br>$\Delta L_{MBMT}$ =0.120<br>$\Delta L_{MBME}$ =0.185 | $\Delta L_{LS}$ =0.325<br>$\Delta L_{MBVSTT}$ =0.325<br>$\Delta L_{MBMT}$ =0.291<br>$\Delta L_{MBME}$ =0.367 | $\Delta L_{LS}$ =0.467<br>$\Delta L_{MBVSTT}$ =0.467<br>$\Delta L_{MBMT}$ =0.451<br>$\Delta L_{MBME}$ =0.502 | $\Delta L_{LS}$ =0.718<br>$\Delta L_{MBVSTT}$ =0.718<br>$\Delta L_{MBMT}$ =0.744<br>$\Delta L_{MBME}$ =0.752 |
| β/Z | 60 | 80 | 92 | 104 | 114 |
| 0.85 | $\Delta L_{LS}$ =0.659<br>$\Delta L_{MBVSTT}$ =0.659<br>$\Delta L_{MBMT}$ =0.681<br>$\Delta L_{MBME}$ =0.662 | $\Delta L_{LS}$ =0.903<br>$\Delta L_{MBVSTT}$ =0.903<br>$\Delta L_{MBMT}$ =0.912<br>$\Delta L_{MBME}$ =0.914 | $\Delta L_{LS}$ =1.040<br>$\Delta L_{MBVSTT}$ =1.040<br>$\Delta L_{MBMT}$ =1.039<br>$\Delta L_{MBME}$ =1.051 | $\Delta L_{LS}$ =1.145<br>$\Delta L_{MBVSTT}$ =1.145<br>$\Delta L_{MBMT}$ =1.157<br>$\Delta L_{MBME}$ =1.150 | $\Delta L_{LS}$ =1.170<br>$\Delta L_{MBVSTT}$ =1.170<br>$\Delta L_{MBMT}$ =1.251<br>$\Delta L_{MBME}$ =1.17 |
| 0.90 | $\Delta L_{LS}$ =0.733<br>$\Delta L_{MBVSTT}$ =0.733<br>$\Delta L_{MBMT}$ =0.762<br>$\Delta L_{MBME}$ =0.736 | $\Delta L_{LS}$ =1.024<br>$\Delta L_{MBVSTT}$ =1.024<br>$\Delta L_{MBMT}$ =1.042<br>$\Delta L_{MBME}$ =1.033 | $\Delta L_{LS}$ =1.196<br>$\Delta L_{MBVSTT}$ =1.196<br>$\Delta L_{MBMT}$ =1.199<br>$\Delta L_{MBME}$ =1.202 | $\Delta L_{LS}$ =1.338<br>$\Delta L_{MBVSTT}$ =1.338<br>$\Delta L_{MBMT}$ =1.346<br>$\Delta L_{MBME}$ =1.343 | $\Delta L_{LS}$ =1.392<br>$\Delta L_{MBVSTT}$ =1.392<br>$\Delta L_{MBMT}$ =1.462<br>$\Delta L_{MBME}$ =1.392 |
| 0.95 | $\Delta L_{LS}$ =0.802<br>$\Delta L_{MBVSTT}$ =0.802<br>$\Delta L_{MBMT}$ =0.838<br>$\Delta L_{MBME}$ =0.804 | $\Delta L_{LS}$ =1.140<br>$\Delta L_{MBVSTT}$ =1.140<br>$\Delta L_{MBMT}$ =1.169<br>$\Delta L_{MBME}$ =1.148 | $\Delta L_{LS}$ =1.345<br>$\Delta L_{MBVSTT}$ =1.345<br>$\Delta L_{MBMT}$ =1.354<br>$\Delta L_{MBME}$ =1.354 | $\Delta L_{LS}$ =1.527<br>$\Delta L_{MBVSTT}$ =1.527<br>$\Delta L_{MBMT}$ =1.529<br>$\Delta L_{MBME}$ =1.534 | $\Delta L_{LS}$ =1.614<br>$\Delta L_{MBVSTT}$ =1.614<br>$\Delta L_{MBMT}$ =1.667<br>$\Delta L_{MBME}$ =1.613 |
| 0.97 | $\Delta L_{LS}$ =0.829<br>$\Delta L_{MBVSTT}$ =0.829<br>$\Delta L_{MBMT}$ =0.867<br>$\Delta L_{MBME}$ =0.831 | $\Delta L_{LS}$ =1.184<br>$\Delta L_{MBVSTT}$ =1.184<br>$\Delta L_{MBMT}$ =1.218<br>$\Delta L_{MBME}$ =1.196 | $\Delta L_{LS}$ =1.404<br>$\Delta L_{MBVSTT}$ =1.404<br>$\Delta L_{MBMT}$ =1.415<br>$\Delta L_{MBME}$ =1.419 | $\Delta L_{LS}$ =1.601<br>$\Delta L_{MBVSTT}$ =1.601<br>$\Delta L_{MBMT}$ =1.600<br>$\Delta L_{MBME}$ =1.723 | $\Delta L_{LS}$ =1.702<br>$\Delta L_{MBVSTT}$ =1.702<br>$\Delta L_{MBMT}$ =1.746<br>$\Delta L_{MBME}$ =1.723 |
| 0.99 | $\Delta L_{LS}$ =0.855<br>$\Delta L_{MBVSTT}$ =0.855<br>$\Delta L_{MBMT}$ =0.896<br>$\Delta L_{MBME}$ =0.889 | $\Delta L_{LS}$ =1.228<br>$\Delta L_{MBVSTT}$ =1.228<br>$\Delta L_{MBMT}$ =1.266<br>$\Delta L_{MBME}$ =1.262 | $\Delta L_{LS}$ =1.461<br>$\Delta L_{MBVSTT}$ =1.461<br>$\Delta L_{MBMT}$ =1.474<br>$\Delta L_{MBME}$ =1.506 | $\Delta L_{LS}$ =1.675<br>$\Delta L_{MBVSTT}$ =1.675<br>$\Delta L_{MBMT}$ =1.671<br>$\Delta L_{MBME}$ =1.719 | $\Delta L_{LS}$ =1.789<br>$\Delta L_{MBVSTT}$ =1.789<br>$\Delta L_{MBMT}$ =1.825<br>$\Delta L_{MBME}$ =1.825 |

[7] The authors of [21, 22] essentially consider one and the same integral (4) [21], proportional to the difference of transport cross sections (8) from [22], which is taken in different ways in these works. Since this integral is the difference of two diverging integrals, the result depends on the method of its finding. As a result, the authors of [21] get only the Mott correction, which must then be summed up with the Bloch correction, while the authors of [22] immediately get this sum.

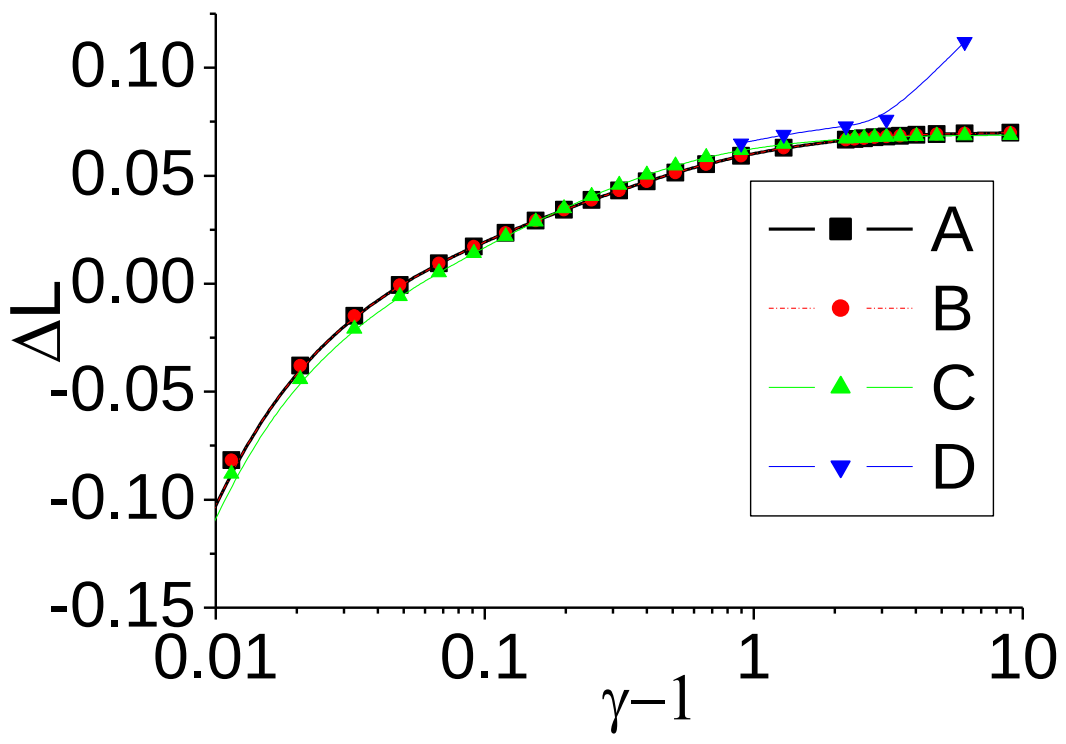

Figure 1.1: $Z = 6$. A: $\Delta L_{\mathrm{LS}}$; B: $\Delta L_{\mathrm{MBVSTT}}$; C: $\Delta L_{\mathrm{MBMT}}$; D: $\Delta L_{\mathrm{MBME}}$.

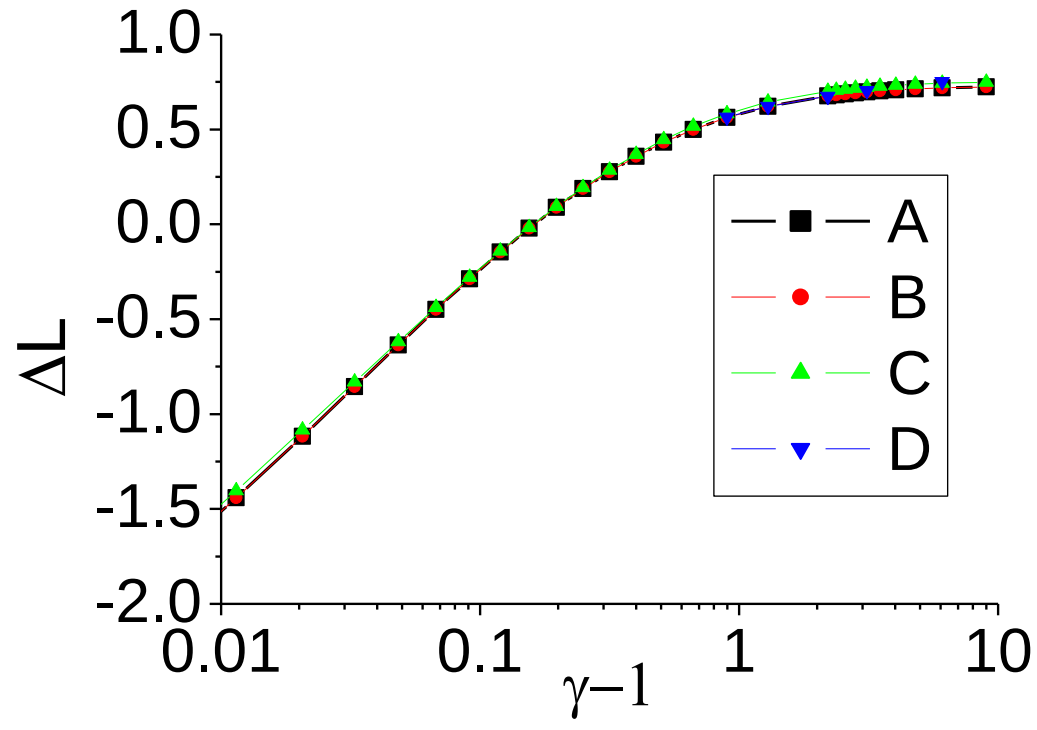

Figure 1.2: $Z = 52$. A: $\Delta L_{\mathrm{LS}}$; B: $\Delta L_{\mathrm{MBVSTT}}$; C: $\Delta L_{\mathrm{MBMT}}$; D: $\Delta L_{\mathrm{MBME}}$.

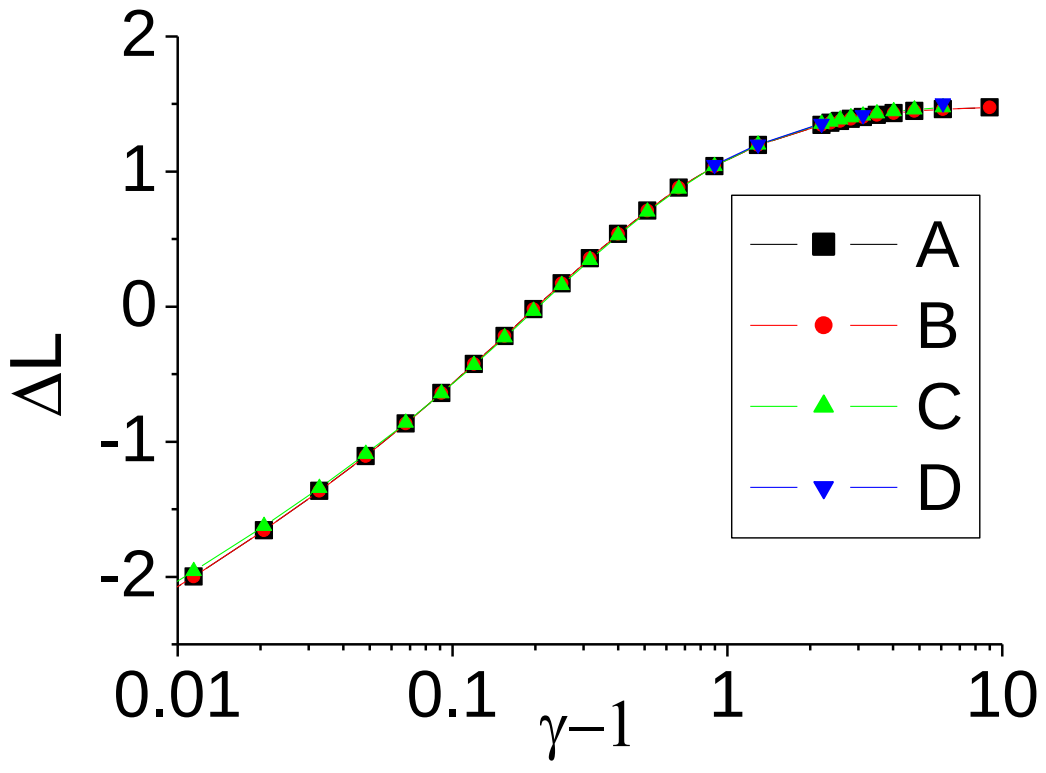

Figure 1.3: $Z = 92$. A: $\Delta L_{\mathrm{LS}}$; B: $\Delta L_{\mathrm{MBVSTT}}$; C: $\Delta L_{\mathrm{MBMT}}$; D: $\Delta L_{\mathrm{MBME}}$.

Figure 1: Lindhard−Sørensen correction (A) in the point nucleus approximation and the Mott−Bloch correction obtained by the VSTT (B) MT (C), and ME (D) methods over the range $0.15 \leq \beta \leq 0.995$ for $Z = 6$ (1.1), 52 (1.2), and 92 (1.3).

For a number of values of $Z$ and $\beta$, there are significant differences in the corrections $\Delta L_{\mathrm{MBVSST}}$ and $\Delta L_{\mathrm{MBME}}$, which was already noted in [31]. The coincidence of the calculation results for $\Delta L_{\mathrm{LS}}$ and $\Delta L_{\mathrm{MBVSST}}$ suggests that these discrepancies are related to the typos or computational errors in [15].

The dependence of the $\Delta L_{\text{LS}}$ correction on the $\gamma$ factor in the Figure 1 exactly corresponds to the same dependence presented in [22].

## 4 Relative difference between the Lindhard−Sørensen and Mott−Bloch corrections

We also evaluated the relative difference $\delta\Delta L$ between the Lindhard−Sørensen and Mott−Bloch corrections,

$$\delta\Delta L = \frac{\Delta L_{MBVSTT} - \Delta L_{LS}}{\Delta L_{LS}} 100\%,$$

as a function of the upper summation limit $N$. Figure 4 shows that the $\delta\Delta L$ value becomes less than 1% already at $N$ = 100 for $Z$ = 118 and $\beta$ = 0.6. For smaller $Z$, the value of $\delta\Delta L$ = 1% is reached even faster. At $N > 600$, the relative error when using $\Delta L_{\text{MBVSST}}$ is less than 0.1%.

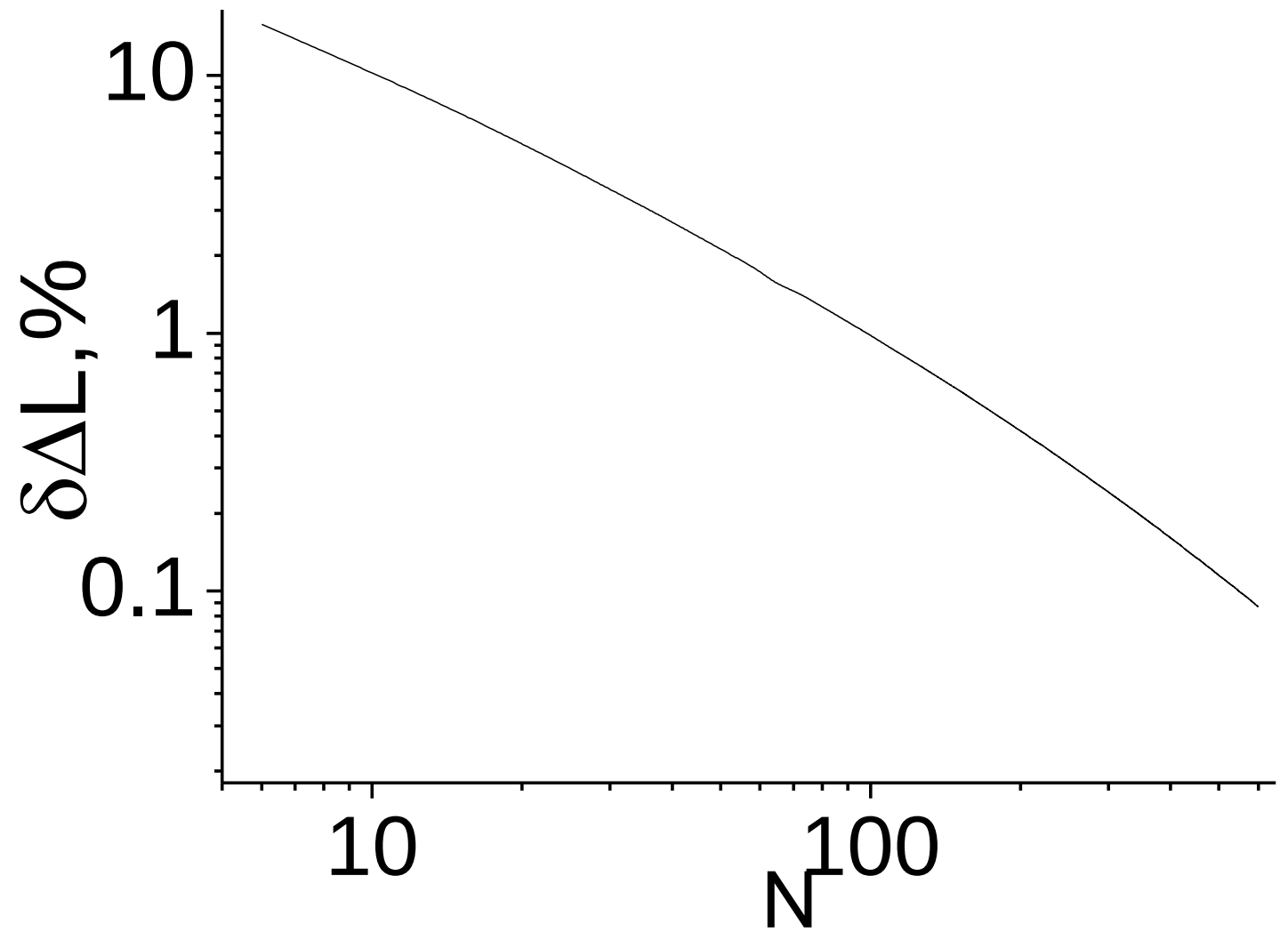

Figure 2**:** Dependence of the relative difference $\delta\Delta L$ between the Lindhard−Sørensen and Mott−Bloch corrections on the upper summation limit $N$ (for $Z$ = 118, $\beta$ = 0.6).

When we represent this difference as a series using (12),

$$\Delta L_{MBVSTT} = \frac{2}{\eta^2}\sum_{k=0}^{\infty}\left[\frac{k(k+1)+\xi^2}{2k+1}\left(\left|F_M^{(k)}\right|^2 - \left|F_Z^{(k)}\right|^2\right) + \frac{k+1}{(k+1)^2+\eta^2} - \frac{1}{k+1}\right],$$

the modulus of its value becomes less than 0.05 % already at $N$ = 65 (Figure 3).

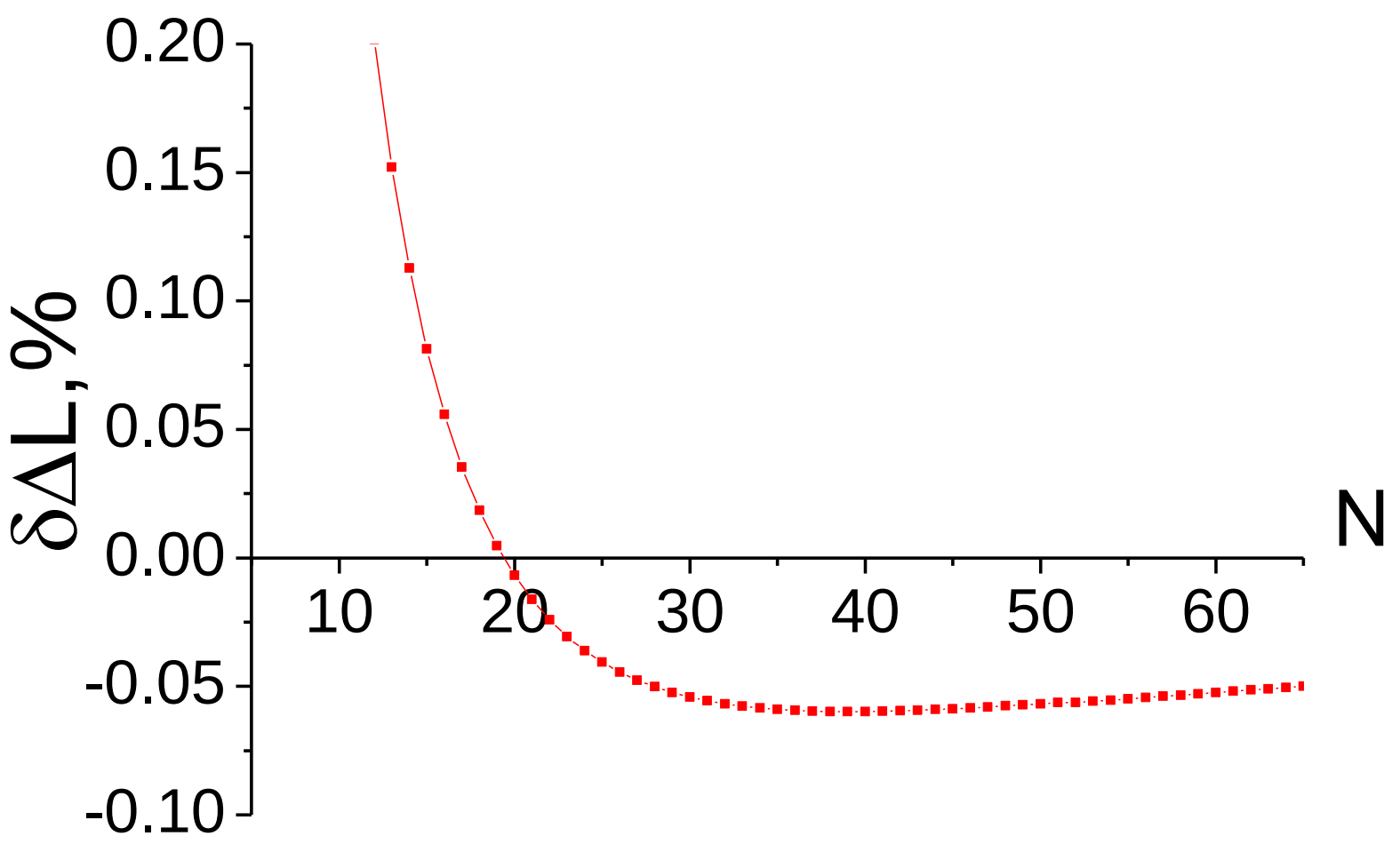

Figure 3: Dependence of the relative difference between the Lindhard−Sørensen and Mott−Bloch corrections on the upper summation limit $N$ (for $Z$ = 118 and $\beta$ = 0.6).

## 5 Mott's correction and difference between the Lindhard−Sørensen and Bloch corrections

We also examined how closely the difference between the Lindhard−Sørensen and Bloch corrections $\Delta L_{LS-B}$ coincides with the Mott correction $\Delta L_M$ calculated by the method [21], which does not use perturbation theory. In calculating the corrections (12) and (9), summation to $k$ = 5000 was performed using the Wolfram Mathematica CAS. Table 2 shows the results of these calculations for uranium ($Z$ = 92). It can be seen excellent agreement between the $\Delta L_{LS-B}$ and $\Delta L_{MVSTT}$ corrections with an accuracy of 6 significant digits. Thus, $\Delta L_{LS-B}$ is close to the exact in $Z\alpha$ correction $\Delta L_M$, and not to its linear approximation (5) as stated in some Refs.

Figure 4 shows the values of these corrections, as well as the Mott correction calculated by the approximate methods of Jackson and McCurthy ($\Delta L_{MJM}$), Morgan and Eby ($\Delta L_{MME}$), and Ahlen ($\Delta L_{MA}$) over the range $0.0500 \le \beta \le 0.9999$ for a number of elements.

Table 2: Difference between the Lindhard−Sørensen correction in the point nucleus approximation and the Bloch correction, $\Delta L_{LS-B}$ (12), as well as the Mott correction (9) obtained by the VSTT method for $Z$ = 92 over the $\beta$ range $0.1 \le \beta \le 0.9$.

| β | 0.1 | 0.2 | 0.3 | 0.4 | 0.5 | 0.6 | 0.7 | 0.8 | 0.9 |
|---|---|---|---|---|---|---|---|---|---|
| $\Delta L_{LS-B}$ | 0.0372735 | 0.139856 | 0.293763 | 0.485402 | 0.703029 | 0.936563 | 0.177409 | 1.418710 | 1.655487 |
| $\Delta L_{MVSTT}$ | 0.0372735 | 0.139856 | 0.293763 | 0.485402 | 0.703029 | 0.936563 | 0.177409 | 1.418710 | 1.655487 |

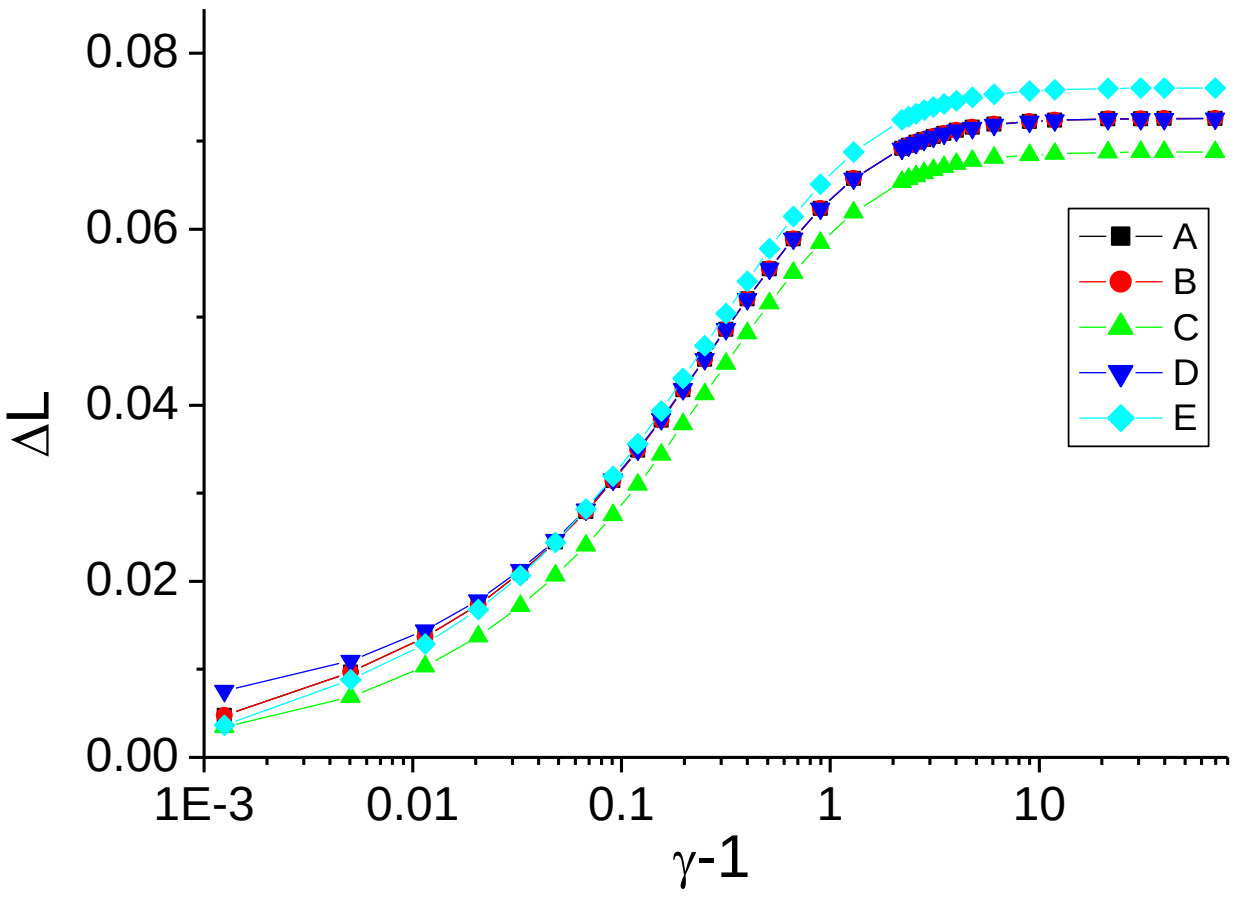

Figure 4.1. $Z = 6$. A: $\Delta L_{\text{LS-B}}$; B: $\Delta L_{\text{MVSTT}}$; C: $\Delta L_{\text{MJM}}$; D: $\Delta L_{\text{MME}}$; E: $\Delta L_{\text{MA}}$.

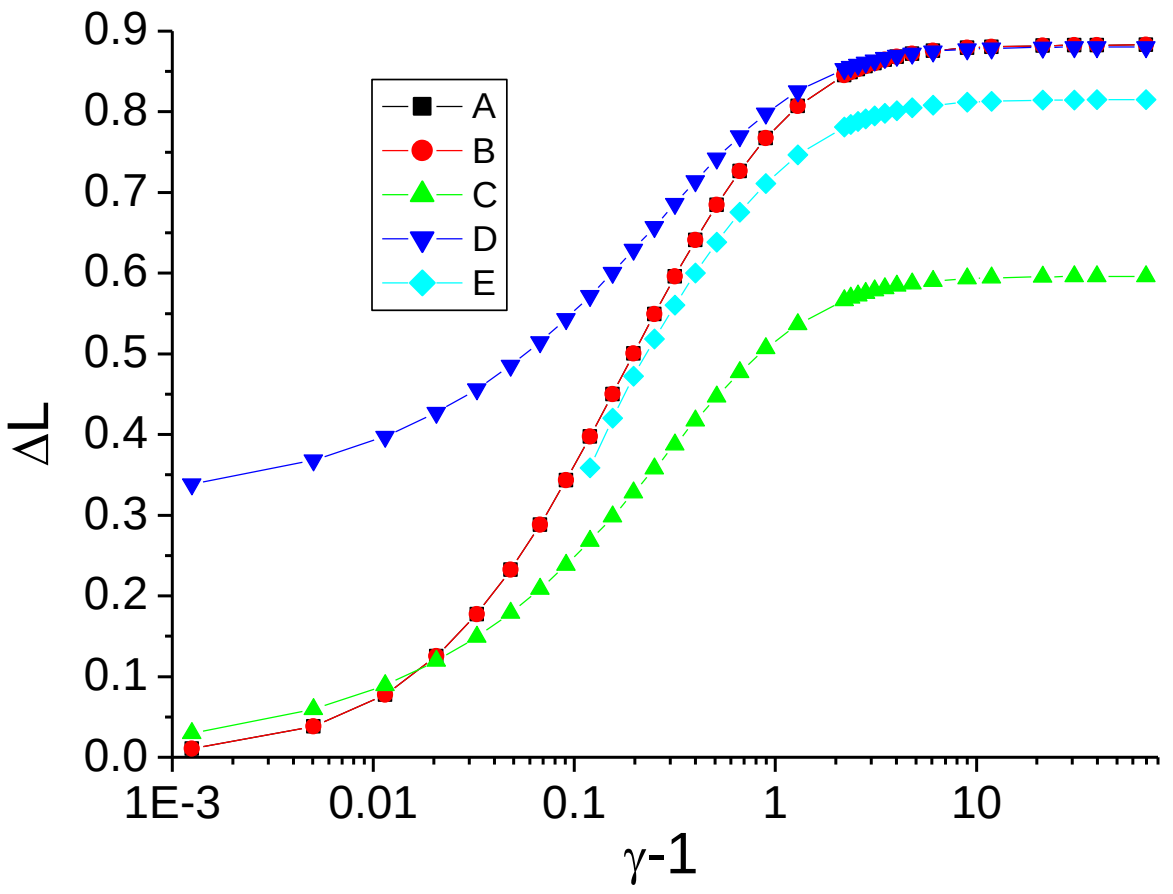

Figure 4.2. $Z = 52$. A: $\Delta L_{\text{LS-B}}$; B: $\Delta L_{\text{MVSTT}}$; C: $\Delta L_{\text{MJM}}$; D: $\Delta L_{\text{MME}}$; E: $\Delta L_{\text{MA}}$.

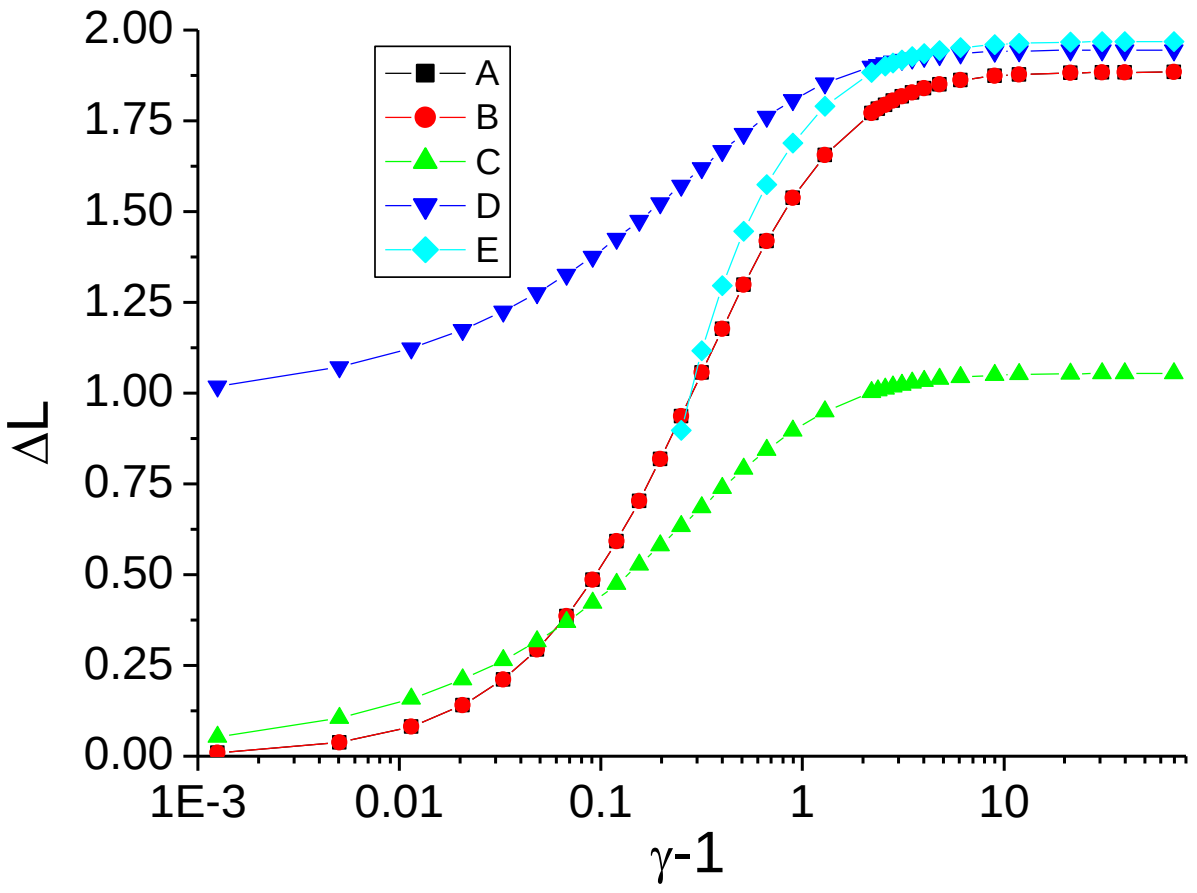

Figure 4.3. $Z = 92$. A: $\Delta L_{\text{LS-B}}$; B: $\Delta L_{\text{MVSTT}}$; C: $\Delta L_{\text{MJM}}$; D: $\Delta L_{\text{MME}}$; E: $\Delta L_{\text{MA}}$.

Figure 4: Difference between Lindhard−Sørensen and Bloch correction (A) in the point nucleus approximation and the Mott correction obtained by the VSTT (B), JM (C), ME (D), and A (E) methods over the range $0.0500 \le \beta \le 0.9999$ for $Z = 6$ (4.1), 52 (4.2), and 92 (4.3).

Figure 4.1 shows that for small $Z$, all approximations give a result close to that obtained in [22]. However, at medium and high values of $Z$, the method of Jackson and McCarthy gives very underestimated values of Mott's correction (Figures 4.2 and 4.3). The method of Morgan and Eby provides the best result for small $Z$. However, as can be seen from Figure 4, this method gives the incorrect behavior of the Mott correction at small $\beta$ values, which is especially noticeable at medium and high $Z$ values. Equation (5) also predicts a nonzero value of Mott's correction when $\beta$ tends to zero. Allen's approximation gives a correction, $\Delta L_{MA,}$ that is less than $\Delta L_{LS-B}$ correction by 8% at $Z$ = 52 (Figure 4.2) and more than $\Delta L_{LS-B}$ by 4% at $Z = 92$ and $\beta = 0.9999$ (Figure 4.3), which is consistent with the conclusions of [21]; in other words, $\Delta L_{MA}$ has uncertain accuracy. At low energies, this approximation leads to non-physical negative values of $\Delta L_{MA}$, according to performed calculations. So, for example, while $\Delta L_{MVSTT} = \Delta L_{LS-B} = 0.125079$ for $Z = 52$ and $\beta = 0.2$, the corresponding $\Delta L_{MA}$ value is $-0.772283$. Thus, the obtained results confirm the conclusion of [21] about the incorrect behavior of some approximate results for Mott's corrections at $\beta \to 0$, as well as about their limited range of applicability and uncertain accuracy.

## 5 Numerical results for stopping power

To compare various methods for calculating stopping power between each other and with experiment, we calculated the quantity $S(E) \equiv -d\,\bar{E}/(\rho dx)$ (1) with the stopping numbers $L_0$ and $L = L_0 + \Delta L$, where the $\Delta L$ means the total Mott−Bloch correction $\Delta L_{MB}$ (4),[8] calculated using formulas (5), (7)-(9), and the Lindhard−Sørensen correction $\Delta L_{LS}$ (10) (Tables 3.1, 3.2, and Figure 5). The results of calculations from [8] in the first-order Born approximation (third column) and based on the Mott exact cross section (seventh column), when they are different from our results, are given in brackets.

Tables 3.1 and 3.2 show that the results obtained by the method [21] are close to those obtained in [8] by integrating the Mott scattering cross section. Since the calculations by the latter method are much simpler, this demonstrated the efficiency of using this method instead of the standard method of integrating the Mott-exact scattering cross section in the case when the lower integration limit tends to zero. Table 3 also demonstrates that the results obtained by the latter method coincide with the results of calculating the stopping power by the method of Lindhard and Sørensen up to the seventh significant digit. It is also obvious from it that the results obtained by these three methods ([8], [21], and [22]) and the Matveev method are consistent with the experimental ones within the experimental error, whereas the Ahlen and Jackson−McCarthy methods give understated values in comparison with the experiment (see Table 3.1 and Figure 5).

[8] The 'Mott−Bloch−Ahlen correction' [25] was calculated according to [32] to ensure its correct comparison with the Mott−Bloch correction calculated by other methods.

Table 3: Electronic stopping power $S(E)$ in MeV cm$^2$ mg$^{-1}$, calculated without $\Delta L$, with the total Mott−Bloch corrections $\Delta L_{MBJM}$, $\Delta L_{MBA}$, $\Delta L_{MBMT}$, and $\Delta L_{MBVSST}$, as well as with the Lindhard−Sørensen correction $\Delta L_{LS}$, in comparison with experimental data from [8].

3.1: Low-*Z* particles.

| Projectile | Target | $S_0$ | $S_{MBJM}$ | $S_{MBA}$ | $S_{MBMT}$ | $S_{MBVSST}$ | $S_{LS}$ | Experiment |
|---|---|---|---|---|---|---|---|---|
| $^{18}_{8}O$ 690 MeV/u ($\beta$=0.819) | Be | 0.125035 | 0.125933 | 0.126061 | 0.126004 | 0.126022 | 0.126022 | 0.125±0.002 |
| | C | 0.137066 | 0.138077 | 0.138220 | 0.138156 | 0.138178 | 0.138178 | 0.138±0.004 |
| | Al | 0.122963 | 0.123937 | 0.124076 | 0.124014 | 0.124035 | 0.124035 | 0.123±0.004 |
| | Pb | 0.082791 | 0.083591 | 0.083705 | 0.083655 | 0.083671 | 0.083671 | 0.084±0.002 |
| $^{40}_{18}Ar$ 985 MeV/u ($\beta$=0.874) | Be | 0.573850 | 0.582735 | 0.585039 | 0.583828 | 0.584732 | 0.584732 | 0.578±0.016 |
| | C | 0.628435 (0.629) | 0.638435 | 0.641029 | 0.639665 | 0.640683 | 0.640683 | 0.640±0.019 |
| | Al | 0.568963 | 0.578608 | 0.581110 | 0.579794 | 0.580776 | 0.580776 | 0.584±0.019 |
| | Cu | 0.494021 | 0.503157 | 0.505526 | 0.504280 | 0.505210 | 0.505210 | 0.494±0.016 |
| | Pb | 0.386315 | 0.394237 | 0.396292 | 0.395211 | 0.396018 | 0.396018 | 0.389±0.012 |

3.2: Medium-*Z* particles .

| Projectile | Target | $S_0$ | $S_{MBJM}$ | $S_{MBA}$ | $S_{MBMT}$ | $S_{MBVSST}$ | $S_{LS}$ | Experiment |
|---|---|---|---|---|---|---|---|---|
| $^{86}_{36}Kr$ 900 MeV/u ($\beta$=0.861) | Be | 2.34572 | 2.40567 | 2.43801 | 2.43794 | 2.43738 (2.438) | 2.43738 | 2.432±0.037 |
| $^{136}_{54}Xe$ 780 MeV/u ($\beta$=0.839) | Be | 5.48721 (5.488) | 5.65418 | 5.70788 | 5.82166 | 5.81012 (5.812) | 5.81012 | 5.861±0.076 |
| | C | 6.01291 (6.014) | 6.20084 | 6.26128 | 6.38934 | 6.37635 (6.378) | 6.37635 | 6.524±0.084 |
| | Al | 5.40984 (5.404) | 5.59110 | 5.64940 | 5.77291 | 5.76038 (5.755) | 5.76038 | 5.806±0.121 |
| | Cu | 4.70236 (4.703) | 4.87404 | 4.92926 | 5.04624 | 5.03438 (5.036) | 5.03438 | 5.077±0.066 |

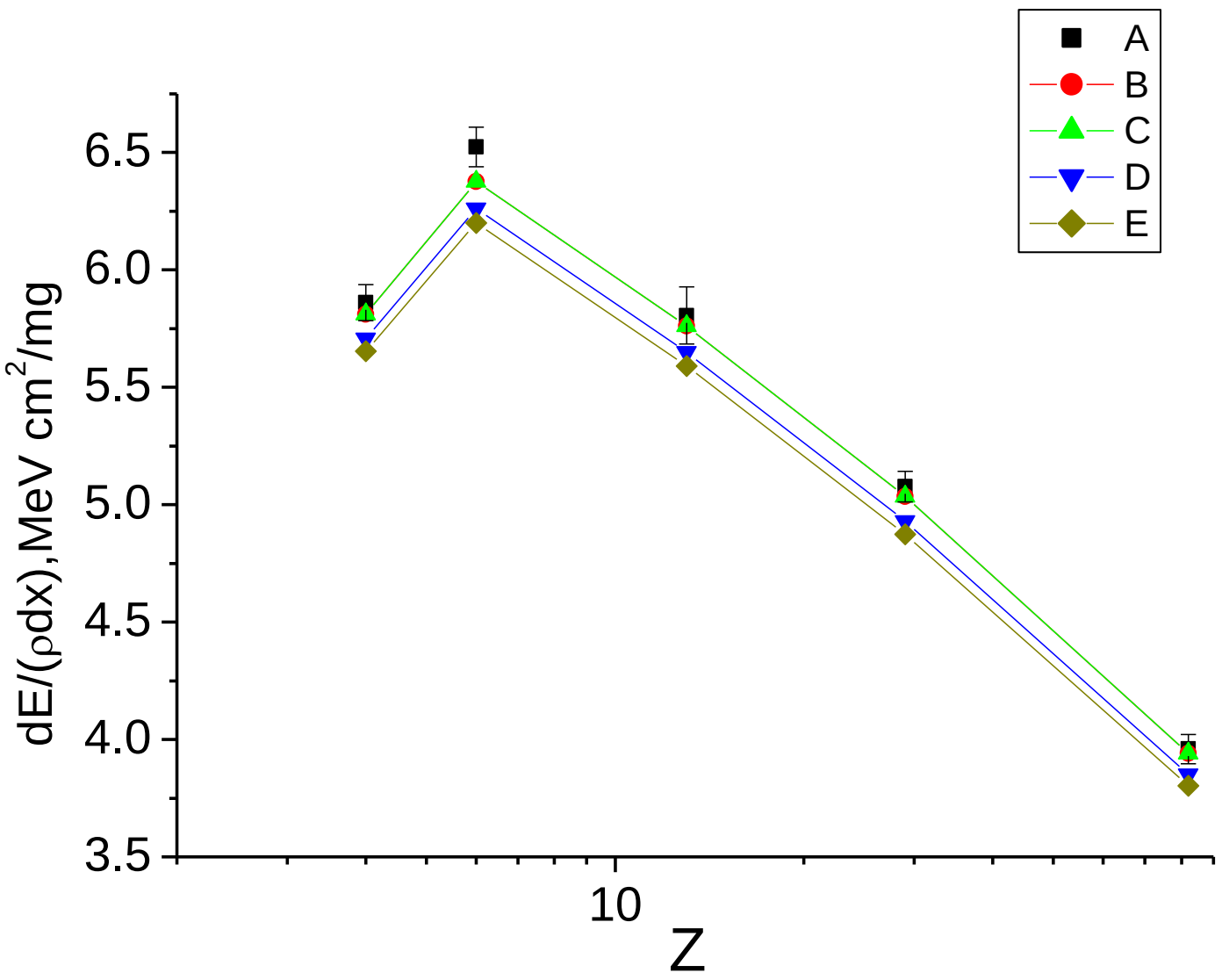

Figure 5: Ionization losses of relativistic ($\beta$ = 0.839) Xe particles in the Be, C, Al, Cu, and Pb targets (left to right): experimental (A) and calculated values with the corrections $\Delta L_{\mathrm{LS}}$ (B), $\Delta L_{\mathrm{MBVSST}}$ (C), $\Delta L_{\mathrm{MBA}}$ (D), and $\Delta L_{\mathrm{MBJM}}$ (E).

The results obtained confirm the conclusions made in [15] that the Bethe formula gives a large error in the computing the ionization losses by heavy ions in solids.

## Summary and conclusions

- In this work, numerical implementation the VSTT method [21] based on the calculation of the Mott exact cross section is given and the preference for using this method instead of the standard method of integrating the Mott cross section is demonstrated for the case when the lower integration limit tends to zero.
- Using the latter result, the Mott correction ($\Delta L_{\mathrm{M}}$) and the total the Mott−Bloch corrections were computed for the ranges of a gamma factor of approximately $1 \lesssim \gamma \lesssim 10$ and the ion nuclear charge number $6 \leq Z \leq 114$.
- The Lindhard−Sørensen corrections in the point nucleus approximation and also the difference between the Lindhard−Sørensen and Bloch corrections ($\Delta L_{LS-B}$) were also calculated in the $\gamma$ and $Z$ ranges under consideration.
- It is shown that the difference between the Lindhard−Sørensen and Bloch corrections and the Mott correction obtained by the exact in $Z\alpha$ VSTT method coincide up to the seventh decimal digit over the range of approximately $1 \lesssim \gamma \lesssim 15$.
- In contrast by the two above-mentioned rigorous methods, the approximate methods have a very limited range of applicability and either (i) give a large difference in the $\Delta L_{\mathrm{M}}$ values (as, for example, the Jackson−McCarthy method in the $\gamma$ range about from 1.01 to 15), or (ii) have an incorrect threshold behavior (e.g. the Morgan−Eby method in the $\gamma$ range from 1 to 2), or (iii) are characterized by an uncertain accuracy (for example, Ahlen's method in the $\gamma$ range about from 1.01 to 15, which also gives non-physical negative values at $\gamma$ less than 1.01) for medium- and high-$Z$ ions. For low-$Z$ ions, these methods give the $\Delta L_{\mathrm{M}}$ values rather close to those obtained by rigorous methods.
- Calculation of the total Mott−Bloch correction ($\Delta L_{MB}$) by the VSTT methods and the Lindhard−Sørensen correction ($\Delta L_{LS}$) over the $\gamma$ and $Z$ ranges $0.01 \leq \gamma - 1 \leq 10$ and $6 \leq Z \leq 114$ gives excellent agreement. The relative difference between these two corrections is less than 0.1% at the upper summation limit $N > 600$.
- We also showed that the results of stopping power calculations obtained by the LS and VSTT methods coincide with each other also up to the seventh significant digit and provide the best agreement with experimental data, while the approximate methods of Ahlen and Jackson−McCarthy give understated values in comparison with the experiment for medium-$Z$ particles ($Z$ = 36, 54).

Thus, we can conclude that at intermediate energies, when a heavy ion can be considered as a point-like particle, both methods, the method based on calculating the Mott exact cross section and the Lindhard−Sørensen method, can be successfully used in electronic stopping calculations for relativistic heavy ions.

## Acknowledgments

The authors express their gratitude to Professor Peter Sigmund (University of Southern Denmark) for his interest in our work.